# Polaritonic Feshbach Resonance

N. Takemura[1], S. Trebaol[1], M. Wouters[2], M. T. Portella-Oberli[1] and B. Deveaud[1]

1) Laboratory of Quantum Optoelectronics, EPFL, CH-1015 Lausanne, Switzerland

2) TQC, University of Antwerp, Universiteislpein 1, 2610 Antwerpen, Belgium

**A Feshbach resonance occurs when the energy of two interacting free particles comes to resonance with a molecular bound state. When approaching this resonance, dramatic changes in the interaction strength between the particles occur. Feshbach resonances have been an essential tool to control the atom interactions, which can even be switched from repulsive to attractive [1-4]. Thanks to Feshbach resonances many effects in ultracold atomic gases could be explored [5, 6]. Here we demonstrate a Feshbach resonance based on polariton spinor interactions that characterize the fundamental interaction process in polariton quantum systems. We show the clear enhancement of attractive interactions and the prompt change to repulsive interaction by tuning the energy of two polaritons with anti-parallel spins across the biexciton bound state energy. A mean field two-channel model quantitatively reproduces the experimental results. This observation paves the way for a new tool to tune the polariton interactions and to move forward into quantum correlated polariton physics.**

A semiconductor microcavity is a unique system where exciton-polaritons emerge from the strong coupling between an exciton and a photon. The demonstration of Bose-Einstein condensation of exciton-polaritons in a semiconductor microcavity [7] has raised a large attention and opened a wide field of research on polariton quantum fluids, such as superfluidity [8], quantum vortices [9], and Bogoliubov dispersion [10-12]. Many more examples could be proposed to highlight the fact that polaritons provide a concrete realization of a bosonic interacting many-body quantum system, complementing the work performed on ultracold atom systems.

Additionally, polaritons exhibit a polarization degree of freedom, with a one-to-one connection to two counter circular polarizations for their photonic part. The different excitonic content of both polarization states, results in asymmetric spinor interactions. Such spinor interactions offer a wide range of effects and a very rich physics to explore in semiconductor microcavities [13-18].

In this work, we present the first demonstration of a Feshbach resonance in a polariton semiconductor microcavity. Feshbach biexcitonic resonant scattering is investigated through spectrally resolved circularly polarized pump-probe spectroscopy on a III-V based microcavity (see method section). To bring the energy of a two-lower polariton state in resonance with the biexciton state we change the cavity exciton detuning (Figure 1(a, c)). We evidence the resonant polariton scattering by probing the anti-parallel spin polariton interactions when scanning the two-polariton energy across the bound biexciton state. We clearly show the enhancement of polariton interactions and the change of their character from attractive to repulsive. Moreover, we observe a decrease of the polariton resonance amplitude when the lower polariton energy is in the vicinity of the biexciton energy. The results are modeled by numerical simulations based on a mean field two-channel model that includes coupling between polaritons and biexcitons as a key ingredient. It is worth mentioning that several works [19-22] have already reported observations of coupling between polaritons and biexcitons states without getting into the regime of the Feshbach resonance reported here.

The expected signatures of the Feshbach resonance phenomenon are twofold. First, a strong variation of the strength and sign of the scattering amplitude and second, a reduction of the free particle density through the coupling with the molecular bound state. Both are sensitive to the energy difference between the free particles and the molecular states. These two energy states refer, in our system, to the state of two anti-parallel spin lower polaritons and the biexciton state, respectively.

We investigate both experimental signatures of the Feshbach resonance. This coherent effect requires working in the coherent regime, therefore we set a zero delay between pump and probe (results from experiments with different pump-probe delays are presented in supplementary material). For a given pump power, we measure through transmitted probe beam the energy shift and the amplitude variation of the lower polariton resonance induced by the presence of the polariton population generated by the pump (Figure 1(b, d, e)). From negative to positive cavity detuning, the energy of the two lower polaritons (2LP, one from the $\sigma_+$ polariton cloud created by the pump, and one $\sigma_-$ coming from the probe) ranges from below to above the molecular biexciton state energy, passing through the resonance (Figure 1(c)). In Figure 2(a), we plot the lower polariton energy shift versus exciton-cavity detuning for a polariton density of $5.1 \times 10^{10}$ polaritons/cm$^2$ generated by the pump. This result clearly shows a dispersive shape characteristic of the resonant scattering. Indeed, in dilute atomic gas systems the Feshbach resonances are evidenced by a dispersive shape of the scattering length at resonance [3]. In these atomic systems, the dispersive shape diverges in the case of magnetic Feshbach resonances with very long molecular lifetime [1, 3]. Here however, a smooth dispersive shape shows up, similar to the case of optical Feshbach resonances with finite molecular lifetime [3, 4]. For polaritons, the finite lifetime of the molecular biexciton state prevents the dispersive shape form diverging at the resonance. Notice that the sign and amplitude of the energy shift are related to the character and strength of the polariton interaction respectively. Our result directly shows that, at resonance, the energy shift switches from redshift to blueshift, demonstrating the drastic modification of the interaction character from attractive to repulsive. This measured shape provides clear evidence for a Feshbach resonance.

In Figure 2(b), we plot the variation of the polariton resonance amplitude in function of the cavity detuning. This result shows the resonant conversion of two polaritons with anti- parallel spins into a biexciton. The simultaneous observations of the change of the polariton interaction character from attractive to repulsive together with the enhanced polariton loss through the coupling with the molecular biexciton state gives direct evidence for the polaritonic Feshbach resonant scattering. The resonance effect shown in Figure 2(a) and (b) is located at an exciton-cavity detuning energy of 0.5 meV, which corresponds to the expected energy range, where the two-lower polariton and the biexciton energies overlap (in Figure 2 - green region). This resonance is located 1.5 meV below the exciton energy in agreement with measured biexciton spectral properties of the same sample previously reported in [20].

To highlight the physics behind our observations, we have performed numerical simulations based on a two-channel model, which includes coupling between polariton and biexciton ($g_{BX}$) in addition to the normal mode coupling between exciton and photon ($\Omega_X$). For our system, it is given by the following Hamiltonian [23, 24]:

$$H = \Omega_X (a_{c\sigma} \psi^+_{X\sigma} + a^+_{c\sigma} \psi_{X\sigma}) + \frac{U_{bg}}{2} (\psi^+_{X\sigma} \psi^+_{X-\sigma} \psi_{X-\sigma} \psi_{X\sigma}) + g_{BX} (\psi^+_{BX} \psi_{X\sigma} \psi_{X-\sigma} + \psi_{BX} \psi^+_{X\sigma} \psi^+_{X-\sigma})$$

Here, $a_{c\sigma}$ and $\psi_{X\sigma}$ are photon and exciton annihilation operators. The polarization state of the photon and the exciton are defined by the spin variable (σ,–σ ) as (↑,↓). The coefficient $g_{BX}$ represents the scattering coupling between two anti-parallel spin polaritons and one biexciton. The underlying physical picture is the coupling of two polaritons to the biexciton bound state through the Coulomb interaction of their excitonic content. A background interaction between anti-parallel spin polaritons is represented by $U_{bg}$. This interaction is found to be attractive (supplementary material). After a mean field approximation, an effective quadratic Hamiltonian H$_{eff}$ for the spin-down probe polariton and the biexciton amplitude is represented by the following 3*3 matrix:

$$H_{eff} = \begin{bmatrix} \psi_{X\downarrow} & a_{c\downarrow} & \psi_B \end{bmatrix}^+ \begin{bmatrix} \varepsilon_X + U_{bg} n_{X\uparrow} - i\frac{\gamma_X}{2} & \Omega_X & g_{BX}\sqrt{n_{X\uparrow}} \\ \Omega_X & \varepsilon_C - i\frac{\gamma_c}{2} & 0 \\ g_{BX}\sqrt{n_{X\uparrow}} & 0 & \varepsilon_B - i\frac{\gamma_B}{2} \end{bmatrix} \begin{bmatrix} \psi_{X\downarrow} \\ a_{c\downarrow} \\ \psi_B \end{bmatrix},$$

where the density $n_{x\uparrow}$ corresponds to the excitonic content of spin up polariton density $n_P$ generated by the pump. $\varepsilon_X$, $\varepsilon_C$, $\gamma_B$, $\gamma_C$ and $\gamma_X$ refer to the exciton energy, cavity photon energy, biexciton linewidth, photon linewidth and exciton homogeneous linewidth, respectively. Note that the polaritonic Feshbach resonance is determined by polariton-biexciton scattering term $g_{BX}$ that describes microscopically how the excitonic content of the two anti-parallel spin polaritons interact resonantly with the biexciton state. By analogy to the Feshbach resonance in atomic system, this term corresponds to the coupling between the open and closed channels [3]. A discussion about the direct coupling between the cavity mode and the biexciton state is presented in Supplementary material. Numerical results give the σ- probe spectra from which we can extract the energy position and the amplitude of polariton resonances for $n_p=0$ and $n_p\neq0$ and therefore their energy shift and amplitude variation.

We extract then the energy shifts and the amplitude variations of the polariton resonance between the calculated $n_p=0$ and $n_p\neq0$ probe transmission spectra as a function of cavity detuning. In Figure 2(a) and (b), the orange lines represent, respectively, the polariton energy shift and amplitude variations obtained by the numerical calculations for $n_P=n_0$. Here, $n_0$ is a normalized polariton population density corresponding to the experimentally extracted polariton density of $5.1\times10^{10}$ polaritons/cm$^2$. The other fitting parameters are summarized in the simulation section. Notice that the calculated results overlap very well with the experimental plots showing a very good agreement between theory and experiment. The plotted black lines correspond to the calculated energy shift and amplitude variations using the same parameters, but without the polariton-biexciton scattering term ($g_{BX}$=0). Here, only the background attractive anti-parallel spin polariton interaction exists. Note that even without biexcitonic effect, the mean field polariton energy shift and amplitude variations depend on the cavity detuning. In Figure 2(a) (black curve), the polariton energy redshift increases according to the forth power of the excitonic Hopfield coefficient [23, 25]. The experimental results however strongly deviate from this behavior when the biexciton resonance is approached. This result evidences the fact that the prominent enhancement of the energy redshift and the change of the sign of the interaction from attractive to weak repulsive, can only be accounted for by the polariton-biexciton coupling, confirming the polaritonic Feshbach resonance.

Concerning the change of the amplitude of the polariton field, there exist two contributions. The first one is an effective positive increase of cavity detuning induced by the polariton energy redshift due to the background attractive interaction [26], which reduces the photonic component of polariton and therefore imply a decrease of the amplitude of the polariton resonance (black line in Figure 2(b)). This contribution alone evidently does not reproduce the experimental results. The second one is the decay of two polaritons with anti-parallel spins into the biexciton channel due to polariton-biexciton coupling. As it is clearly shown in Figure 2(b) (orange curve), to reproduce accurately the large decrease of the polariton resonance amplitude the polariton *spin up* + polariton *spin down* → biexciton scattering process should be introduced in the model.

We plot in Figure 2(c) interaction and absorption ratios extracted from Figures 2(a) and (b) respectively. Those ratios result from the total contribution (orange curves) divided by the background contribution (black curves). This two ratios evidence the usual behavior of a scattering resonance. The absorption ratio (curve b) behaves as the expected scattering resonance profile due to the two-body loss process of lower polaritons when crossing the molecular state. The interaction ratio (curve a) displays a dispersive shape of the scattering length showing a large variation in term of sign and amplitude of the interaction strength crossing the biexciton resonance. Analogous observations are reported by Theis et al. [4] in $^{87}$Rb condensate in optical Feshbach resonance.

This model, in which we consider the interaction of the two anti-parallel spin polaritons with the molecular biexciton state, reproduces the experimental results very well. This is the resonant interaction term that governs a Feshbach resonance. In the present model, it is important to note that the coupling between polaritons and biexcitons, the coupling term $g_{BX}\sqrt{n_X}$, depends not only on the constant $g_{BX}$ but also on the excitonic content of polariton population $\sqrt{n_X}$. At low pump power, the pump polariton density is small and thus this coupling term is weak. In this perturbative regime, the comparison between simulations and experiments shows excellent qualitative and quantitative agreements demonstrating unambiguously the polaritonic Feshbach resonance.

We now turn to the investigation of the polaritonic Feshbach resonance in a system with a larger polariton population. We perform pump and probe measurements as a function of the cavity detuning using higher pump intensities in order to increase the polariton population. As the pump power is increased, a new resonance shows up at higher energy than the polariton resonance (Fig. 1(e)). In Figure 3 we plot, for a polariton density of 1.6×10$^{11}$ polaritons/cm$^2$ generated by the pump, the energy shift of both polariton resonances and the amplitude variation of the lower energy polariton resonance as a function of the cavity detuning.

When the cavity detuning brings the two-polariton state energy close to the biexciton bound state energy then both resonances show an anti-crossing behavior (Figure 3(a)). The associated energy shifts obtained from the numerical simulations using the same parameters as for the lower power regime but now with $n_p$=2.44$n_0$ appear as full and dashed orange lines for the lower and higher energy resonances respectively. The numerical results show remarkable quantitative agreement with the experimental ones. We also display the numerical result obtained without considering the polariton-biexciton coupling ($g_{BX}$=0) as solid black line, which clearly deviates from the experimental data. In this high density regime, the appearance of a new resonance and the anti-crossing behavior evidences the strong polariton-biexciton coupling. It is important to note that, to reach this strong coupling regime, the polariton-biexciton coupling energy $g_{BX}\sqrt{n_X}$ should be comparable to the linewidths of the involved states. In our system, the polariton-biexciton coupling strength is directly controlled by the density of polaritons through the pump intensity. Then, for the pump intensity used in Figure 3, the polariton-biexciton coupling energy is of the same order than the biexciton linewidth. Indeed, in this regime, the polaritonic Feshbach resonance signs an anti-crossing between these two states in addition to the changing of the polariton interaction.

Moreover, as the spin up polariton population is increased, the two spin-up spin-down polariton correlation is enhanced, which favors the biexciton creation. In Figure 3(b), we plotted the amplitude variations of the lower energy resonance displayed in Figure 3(a). The enhancement of the amplitude variation with pump intensity and thus the increase of the losses with pump polariton population is evidenced. This supports the enhancement of the biexciton formation for the larger polariton population. The very good agreement between experimental and theoretical values corroborates our interpretation as a polaritonic Feshbach resonance.

All these results together demonstrate two different regimes of polaritonic Feshbach resonance. First, in the low density regime, where polaritons and biexcitons show small interactions, the polariton-biexciton coupling energy cannot overcome the damping rate of the biexcitons $2g_{BX}\sqrt{n_X} < \gamma_B$. Therefore, in this weak interaction regime, the manifestation of the Feshbach resonance shows up a dispersive shape. On the contrary, in the strong interaction regime $2g_{BX}\sqrt{n_X} \sim \gamma_B$, polaritons and biexcitons states manifest an anticrossing. The splitting energy, proportional to twice the polariton-biexciton coupling term, increases for much higher pump intensities (not shown). As a result, the polariton population governs the polariton-biexciton coupling strength, which allows for the polaritonic Feshbach resonance to be tuned from the weakly to strongly interacting regime.

**Methods**

Our study is performed on high quality III-V microcavity. A single 8nm $In_{0.04}Ga_{0.96}As$ quantum well is sandwiched between a pair of GaAs/AlAs distributed Bragg-reflectors (DBRs). The exciton energy of the quantum well is 1.4866 eV and the Rabi splitting at zero detuning between the cavity and exciton energy is 3.26 meV [27]. The measurements are performed at a temperature around 4 K. We employ a counter-circular polarization configuration in order to investigate the polariton-polariton interaction with anti-parallel spins. To demonstrate the Feshbach resonance on the spinor polariton scattering, we carry out polarization resolved pump-probe spectroscopy in transmission. We employ an heterodyne measurement technique [12, 28], which dramatically increases the signal to noise ratio. It allows to resolve small probe beam energy shifts and to use a degenerate beam configuration. Figure 1(b-c) shows the schematics of our pump-probe experiment. The sample is resonantly excited at k=0 with a σ+ circularly polarized pump pulse to generate a spin-up polariton population and probed at k=0 using a counter-circularly polarized probe pulse. Since the cavity spacer layer is wedged, we tune the resonance energy of the cavity by moving the laser spot over the sample. The biexciton binding energy $E_{bix}$ is of the same order as the Rabi splitting. It is then straightforward to bring the energy of a two-lower-polariton state in resonance with the biexciton state by changing the cavity-exciton detuning (Figure 1(a, c)). We repeat the experiment for different cavity detunings and different pump powers.

The laser source is a 125 fs pulse with a repetition rate of 80 MHz, which is separated into three pulses: pump, probe and reference. The pump spot size (2200 $\mu m^2$) is larger than probe's size (250 $\mu m^2$) to ensure the study of a uniform pump polariton density. We fixed the probe intensity about 3 times smaller than the pump intensity in the low density regime (Figure 2). Then, increasing the pump intensity, the pump/probe density ratio is even larger. The center frequency of the laser is set between the lower and upper polaritons. In our measurement we varied the LP energy only within 2 meV. Since the laser spectrum is about 14.6 meV wide, the laser intensity is considered to be constant for all laser detunings. Both probe and reference beams are detected by a spectrometer and the spectrum of the probe pulse is numerically reconstructed.

**Simulation**

The biexciton energy is determined as $\varepsilon_B = \varepsilon_X - E_{bix}/2$, with $E_{bix}$, the biexciton binding energy. For the numerical calculation of Figure 2 and 3 we use the following parameters: the biexciton binding energy $E_{bix}/2$ = 1.5 meV, a biexciton linewidth $\gamma_B$ = 1.1 meV. For the polariton-biexciton coupling, the exciton-exciton scattering term is set to $g_{BX} = 0.36 \frac{meV}{\sqrt{n_0}}$. The background interaction is $U_{bg} = -0.18 \frac{meV}{n_0}$, where $n_0$ is the normalized density of spin up polaritons. Other parameters are photon linewidth $\gamma_C$ = 0.3 meV, exciton linewidth $\gamma_X$ = 0.6 meV and Rabi coupling $\Omega_X$ =1.63 meV. Note that same parameter values are used for all experiments, except the polariton density $n_P$ which is adjusted when the pump density is changed. The excitonic density $n_X$ is expressed in function of the polariton density $n_P$ through the expression:

$$n_X = \frac{n_P}{2}\left(1 + \frac{\delta}{\sqrt{\delta^2 + 4\Omega_X^2}}\right), \tag{S1}$$

where $\delta$ is the exciton-cavity detuning. Polariton density is estimated through the density of the excitonic component. In the higher density experiment, we observe a strong to weak coupling transition at the cavity detuning 0.87 meV. This means that, at this cavity detuning, the excitonic component of polaritons reaches the exciton saturation density $n_{Xsat}$, which is given by $n_{Xsat} = 7/(16\pi a_{2D}^2) \approx 1 \times 10^{11}$ cm$^2$ [29] for a Bohr radius $a_{2D}$=12 nm in a single InGaAs quantum well [30]. The polariton density can be then estimated using (S1). In order to facilitate the extraction of the new upper resonance energy position in the large polariton population experiment (orange dashed line figure 3), we force in the model $\gamma_{bx}$=0 to enhance the peak transmission. Notice that this procedure does not affect the resonance energy positions.

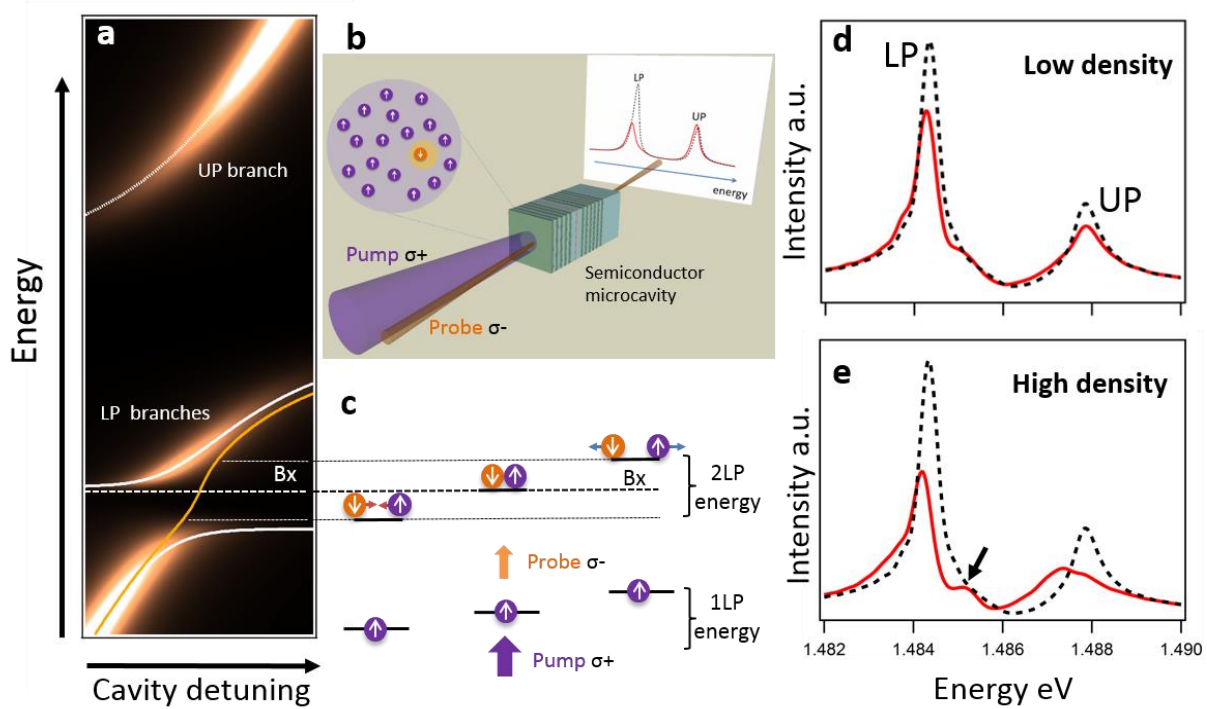

**Figure 1 Experimental scheme used for the polaritonic Feshbach resonance (a)** The spin down polariton energy varies in function of the cavity detuning in presence of a spin up polariton population as: In the perturbative regime, the lower polariton resonance shows a dispersive shape (orange line) around the crossing point with the biexciton energy state (bold dashed line). At higher polariton spin up densities, the lower polariton branch splits in two at the biexciton crossing energy (solid white lines). This effect results from the strong coupling with the biexciton resonance. UP and LP branches stand for the upper and lower polariton resonances respectively. **(b)** A pump beam creates a gas of spin up polaritons (violet) and the probe introduces few spin down polaritons

(orange) which interact with the gas. **Inset (b)** Transmission spectrum of the probe σ-: without (black line) and with (red line) the presence of a spin up polariton gaz. **(c)** Two-polariton energy with respect to the biexciton state determines the interaction configuration: attractive (red arrows) or repulsive (blue arrows). The resulting polariton energy is depicted on Figure 1(a). **(d, e)** Pump-probe spectra in function of the polariton density. For a cavity detuning of -1.1 meV, **(d)** $4.5 \times 10^{10}$ /cm$^2$ and **(e)** $1.5 \times 10^{11}$ /cm$^2$. On Figure **(e)**, the black arrow points out the appearance of a new resonance resulting from the strong coupling between LP branch and biexciton.

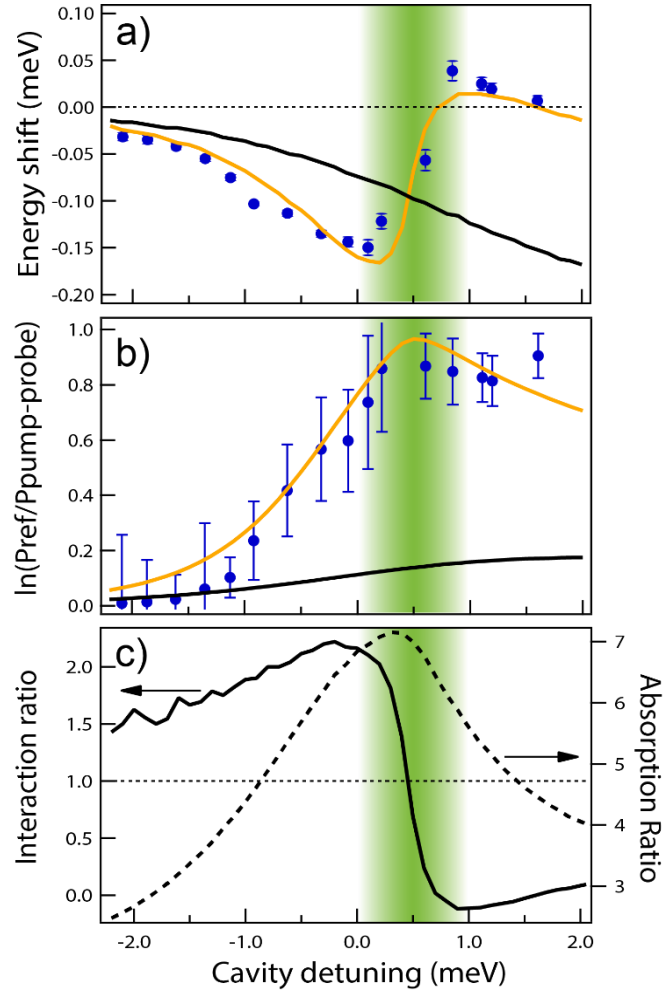

**Figure 2**. **Polaritonic Feshbach resonance: experimental manifestations (a)** Energy shifts of the pump-probe spectrum and **(b)** $ln(P_{ref}/P_{pump-probe})$ as a function of cavity detuning. $P_{ref}$ and $P_{pump-probe}$ respectively represent the amplitude of lower polariton resonance without and with pump excitation. Blue circles are experimental results, orange and black lines stand for numerical simulations with and without polariton-biexciton coupling respectively. **(c)** Interaction (curve a) and absorption (curve b) ratios in function of the cavity detuning. They result from the ratio between the orange and black lines extracted from Figure 2(a) and (b) respectively. Fitting parameters are summarized in simulation section. The fitted density of polaritons is $n_p=n_0$. The green region corresponds to the
13

energy range where the two-polariton energy crosses the biexciton state energy: $E_{LP\uparrow} + E_{LP\downarrow} \leftrightarrow E_{BX}$ (Figure 1(c)). Error bars represent $3\sigma$ with $\sigma$ the standard deviation.

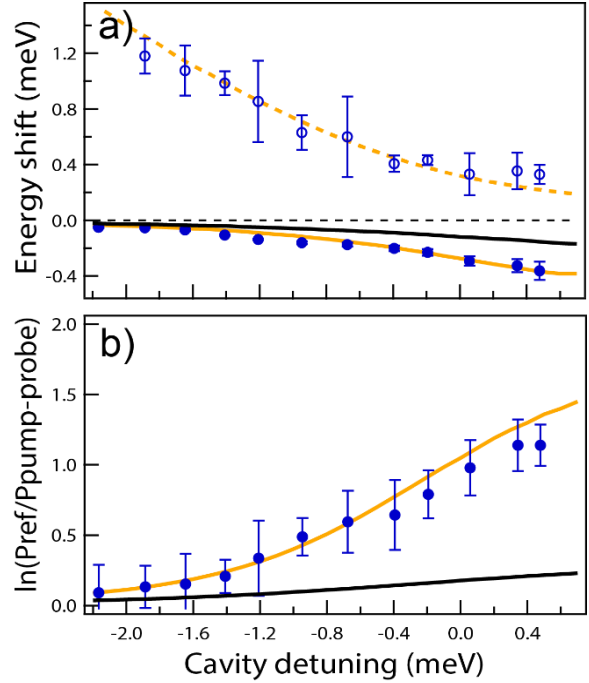

**Figure 3**. **Higher density polaritonic Feshbach effect.** **(a)** Energy shift of the pump-probe spectrum as a function of cavity detuning for a polariton pump density of $1.6\times10^{11}$ polaritons/cm$^2$. For higher pump intensities, measurements are limited below 0.5 meV cavity detuning to avoid a strong to weak coupling transition which render the polariton picture no more appropriate. **(b)** *ln*($P_{ref}$/$P_{pump-probe}$) for the lowest polariton branch in function of the cavity detunings. Orange and Black lines respectively represent numerical simulations with and without polariton-biexciton coupling obtained with a density $n_p=2.44n_0$. Error bars represent $3\sigma$ with $\sigma$ the standard deviation.

**Acknowledgement**

We thank prof. I. Carusotto for fruitful discussions.

The present work is supported by the Swiss National Science Foundation under project N°135003, the Quantum Photonics National Center of Competence in Research N°115509 and by the European Research Council under project Polaritonics contract N°219120. The polatom network is also acknowledged.

**Contributions**

N.T. and S.T. contributed to the experiment, N.T. performed the measurements and the data processing. M.W. developed the theoretical model. N.T. and S.T. did the simulations. N.T., S.T. and M.T.P.O. performed data analysis. S.T. and M.T.P.O. wrote the paper. M.T.P.O. and B.D. supervised the whole project. All authors contributed to numerous discussions and revised the manuscript.

# Supplementary material

# Polaritonic Feshbach Resonance

N. Takemura[1,], S. Trebaol[1], M. Wouters[2], M. T. Portella-Oberli[1] and B. Deveaud[1]

*1) Laboratory of Quantum Optoelectronics, EPFL, CH-1015 Lausanne, Switzerland*

*2) TQC, University of Antwerp, Universiteislpein 1, 2610 Antwerpen, Belgium*

## I The effects of decoherence on polariton Feshbach resonance

### Feshbach resonance and pump-probe delay dependence

In order to show that the polaritonic Feshbach resonance only persists during the polariton coherence time and disappears with the decoherence of the system, we performed the experiments as a function of the delay between pump and probe pulses. Positive (negative) delay means that the pump (probe) arrives first.

In Figure S1 below we show the energy shift (upper panel) and absorption (lower panel) of the polariton resonance as a function of cavity detuning for different delays. The results obtained at zero, negative and positive delays are shown together to highlight the disappearance of the Feshbach resonance with decoherence at positive delay and its gradual decrease in coherent regime at negative delay.

Let us now explain in detail our experiments and our findings:
In the negative delay configuration, the pump-probe signal only exists for delays within the coherence time of the system. Then the pump-probe signal decays together with coherence and therefore with delay. It is important to point out the significance of the experiments at negative delays, in which we only probe the coherent effects.

We plot in Figure S1a the energy shift of the polariton resonance as a function of cavity detuning for zero, -1.5 ps and -3 ps negative delays. The resonance behaviour clearly appears for zero delay, as the result already presented in Figure 2a in the manuscript. As expected the amplitude of the shifts decreases with the increase of the negative delay. In Figure S1b, we plot the absorption strength of the polariton resonance as function of cavity detuning at zero, -1.5 ps and -3 ps negative delays. Here also we clearly see the decrease of the signal when the delay is increased. However it is important to point out that the resonance features (although decaying) are always present.

With these experiments we demonstrate that the polaritonic Feshbach resonance persists during the coherence time of the polariton system.

At positive delays the pump pulse arrives first and the pump-probe signal exists during the lifetime of the population even if the system loses its coherence. Therefore, in this configuration, the pump-probe signal exhibit features related to coherent and incoherent effects.

We plot in Figure S1c the polariton resonance energy shift and in Figure S1d the absorption as a function of cavity detuning obtained at zero, 1.5 ps and 3 ps positive delays. We show the polaritonic Feshbach resonance features at zero delay in which polaritons are in full coherent regime.

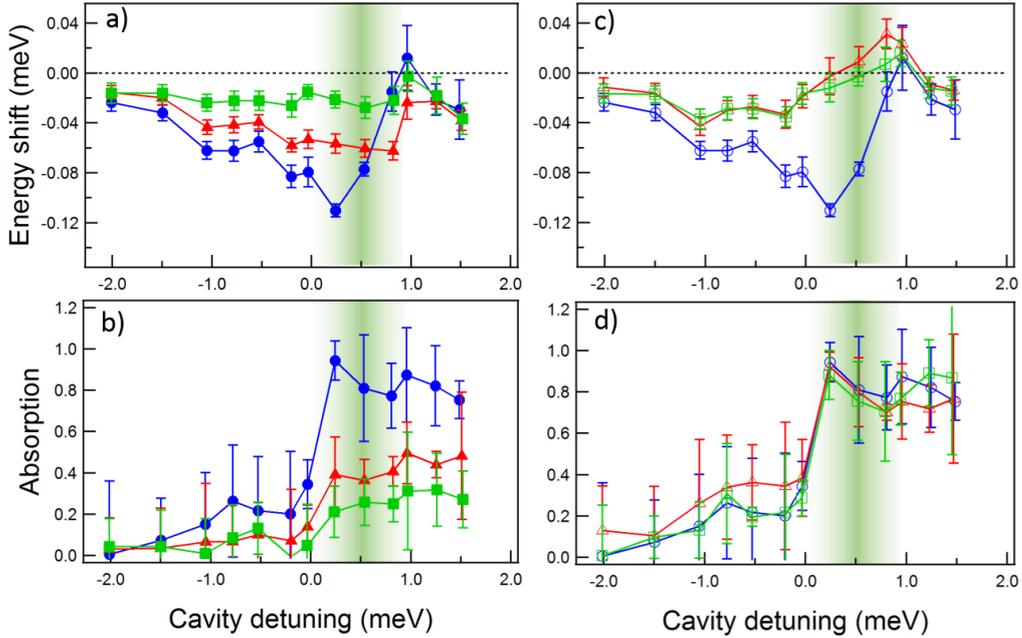

**Figure S1** Energy shifts (upper panel) and absorption $ln(P_{ref}/P_{pump-probe})$ (lower panel) of the lower polariton resonance as a function of cavity detuning for different pump-probe delays. Zero, 1.5 ps and 3ps negative (positive) delays correspond respectively to circle, triangle and square full (empty) markers. $P_{ref}$ and $P_{pump-probe}$ respectively represent the amplitude of lower polariton resonance without and with pump excitation. Feshbach resonance in full coherent regime at zero delay. (a and b) Negative pump-probe delays and the persistence of Feshbach resonance in coherent regime: The signal decreases with coherence and therefore with delay. (c and d) Positive pump-probe delays and the disappearance of Feshbach resonance in incoherent regime: The energy shift resonance feature disappears with decoherence therefore with delay. However the absorption persists in incoherent regime due to the presence of polariton population.

Upon increasing the delay, incoherent effects take place bringing a destruction of the resonance feature. Indeed the dispersive shape energy shift, characteristic of the Feshbach resonance is definitely washed out. Moreover, contrary to the results at negative delays the absorption variation of the polariton resonance persists for positive delays due to the presence of the polariton population.

As a conclusion, measurements performed at negative and positive delays give an unambiguous proof of Feshbach resonance as a coherent effect:

(i)    The Feshbach resonance characteristic features both for the energy shift and the amplitude of the absorption evolve with coherence at negative delay.

(ii)    Decoherence kills Feshbach resonance. This is clearly demonstrated with the disappearance of the resonance feature in the energy shift for positive delays. Because a real population is created at positive delays, the absorption persists manifesting the presence of incoherent population effects.

Both results together evidence that incoherence prevents polaritonic Feshbach resonance.

# II Background interactions of anti-parallel spin polaritons

There are several possible interaction mechanisms that may contribute to the background interactions. We can mention indirect scattering through dark exciton states [26, 31], and also through excited exciton states [32], exciton-exciton scattering continuum correlations [33], which are attractive. Repulsive interactions such as Van-der-Waals and electrostatic interactions [26] could also contribute. However it is very difficult to determine their relative contributions in the overall energy shift as shown by the different models that can be found in the literature [26, 31, 32, 33, and references therein].

We use a phenomenological model in which the background interaction and the biexciton-exciton interaction are two independent constants. Our experimental observations (energy shift and absorption) can only be reproduced using a background interaction constant with a fixed negative value without changing other parameters except the density of polaritons (Figure 2 and 3 in the paper). In addition, neither null nor positive background constant allow to reproduce our data even by trying to adjust the value of the biexciton-exciton coupling $g_{BX}$ constant.

Here background interaction and biexciton-exciton interaction play a very different role to the Feshbach effect. The biexciton-exciton interaction allows to reproduce the two main effects of the Feshbach resonance namely the strong absorption and the change of the magnitude and of the sign of the energy shift (see Figure S2). The background interaction only acts as an offset of the energy shift – which is negative - as illustrated on Figure S2.

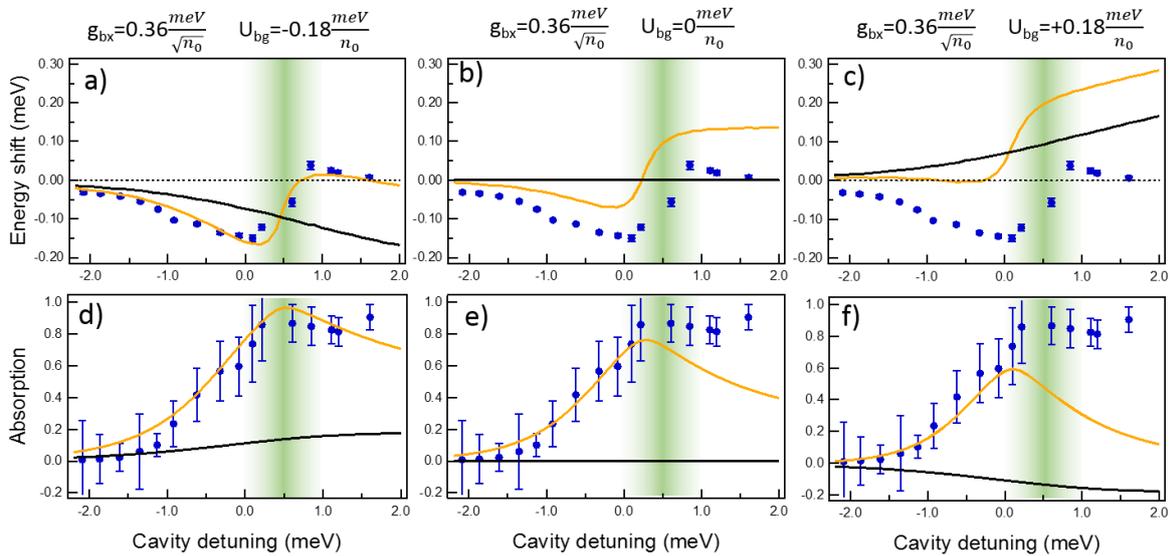

**Figure S2 Effect of the background interaction constant as a fitting parameter on data shown in Figure 2 (in the paper).** *Upper panel* Energy shifts of the pump-probe spectrum and *lower panel*, $ln(P_{ref}/P_{pump\text{-}probe})$ as a function of cavity detuning. Blue circles are experimental results, orange and black lines stand for numerical simulations with and without polariton-biexciton coupling respectively. (a,d), (b,e) and (c,f) show the effect of different values of $U_{bg}$ on the orange fitting curve.

## III Origin of the coupling term in the Feshbach resonance

Different transitions between polariton and biexciton could be observed through pump/probe experiments. First, the direct transition from a polariton to the biexciton state as reported by Borri *et al.* [19]. Second, the direct coupling between an exciton and a cavity photon to reach the biexciton state. This process is at the origin of the oscillator strength transfer from excitons to biexcitons reported in [20].

Setting the two anti-parallel polariton spin state in resonance with the biexciton gives rise to the polaritonic Feshbach resonance. Our Feshbach resonance observation is defined in the framework of the "bipolariton model" [23, 24]. Note that in our model, the polariton-biexciton coupling constant $g_{BX}$ describes the anti-parallel spin exciton+exciton coupling to biexciton. This effect is induced by the Coulomb interaction of excitons. The Feshbach resonance could also be described in the framework of another model the "giant oscillator strength model" [24, 18]. In this model the coupling term is $\Omega_B$, which refers to the photon+exciton coupling to biexciton.

To highlight all pathways allowing to couple to biexcitons, we have included in our "bipolariton model", in addition to the $g_{BX}$ coupling, the coupling term $\Omega_B$. This model Hamiltonian is then rewritten as:

$$H = \Omega_X(a_{c\downarrow}\psi_{X\downarrow}^+ + a_{c\downarrow}^+\psi_{X\downarrow}) + \Omega_B(a_{c\downarrow}\psi_{X\uparrow}\psi_{BX}^+ + a_{c\downarrow}^+\psi_{X\uparrow}^+\psi_{BX} + a_{c\uparrow}\psi_{X\downarrow}\psi_{BX}^+ + a_{c\uparrow}^+\psi_{X\downarrow}^+\psi_{BX}) + \frac{U_{bg}}{2}\left(\psi_{X\uparrow}^+\psi_{X\downarrow}^+\psi_{X\downarrow}\psi_{X\uparrow}\right) + g_{BX}(e^{-i\theta}\psi_{BX}^+\psi_{X\uparrow}\psi_{X\downarrow} + e^{i\theta}\psi_{BX}\psi_{X\uparrow}^+\psi_{X\downarrow}^+)$$

In terms of the 3*3 matrix, after mean field approximation the Hamiltonian is the following:

$$H_{eff} = \begin{bmatrix}\psi_{X\downarrow} & a_{c\downarrow} & \psi_B\end{bmatrix}^+ \begin{bmatrix} \varepsilon_X + U_{bg}n_{X\uparrow} - i\frac{\gamma_X}{2} & \Omega_X & g_{BX}\exp(i\theta)\sqrt{n_{X\uparrow}} - \Omega_B\sqrt{n_{c\uparrow}} \\ \Omega_X & \varepsilon_C - i\frac{\gamma_c}{2} & \Omega_B\sqrt{n_{X\uparrow}} \\ g_{BX}\exp(-i\theta)\sqrt{n_{X\uparrow}} - \Omega_B\sqrt{n_{c\uparrow}} & \Omega_B\sqrt{n_{X\uparrow}} & \varepsilon_B \end{bmatrix} \begin{bmatrix}\psi_{X\downarrow} \\ a_{c\downarrow} \\ \psi_B\end{bmatrix}$$

Here $n_{C\uparrow}$ ($n_{X\uparrow}$) corresponds to the photonic (excitonic) density generated by the pump beam. In addition, we have introduced a phase term between the two couplings $g_{BX}$ and $\Omega_B$ required to generalize the model.

Let us now detail the respective significance of the couplings $g_{BX}$ and $\Omega_B$ in the Feshbach resonance process.

First, in Figure S3a and b (Figure 2a and b in the paper), we show that the experimental results are very well reproduced by our model, when we set $\Omega_B=0$ and consider the $g_{BX}$ coupling as the unique active mechanism in the polaritonic Feshbach resonance. Second, in Figure S3c and d, we now set $g_{BX}=0$ and then we present the results of model when the $\Omega_B$ coupling term is the only responsible for the effect. This model also reproduces the experimental results reasonably well at resonance, but the shape of the absorption feature deviates for both very negative and positive detuning. These results enlighten that indeed both terms are giving rise to a Feshbach resonance, however our experimental observation favors the Coulomb term $g_{BX}$ coupling over the giant oscillator coupling.

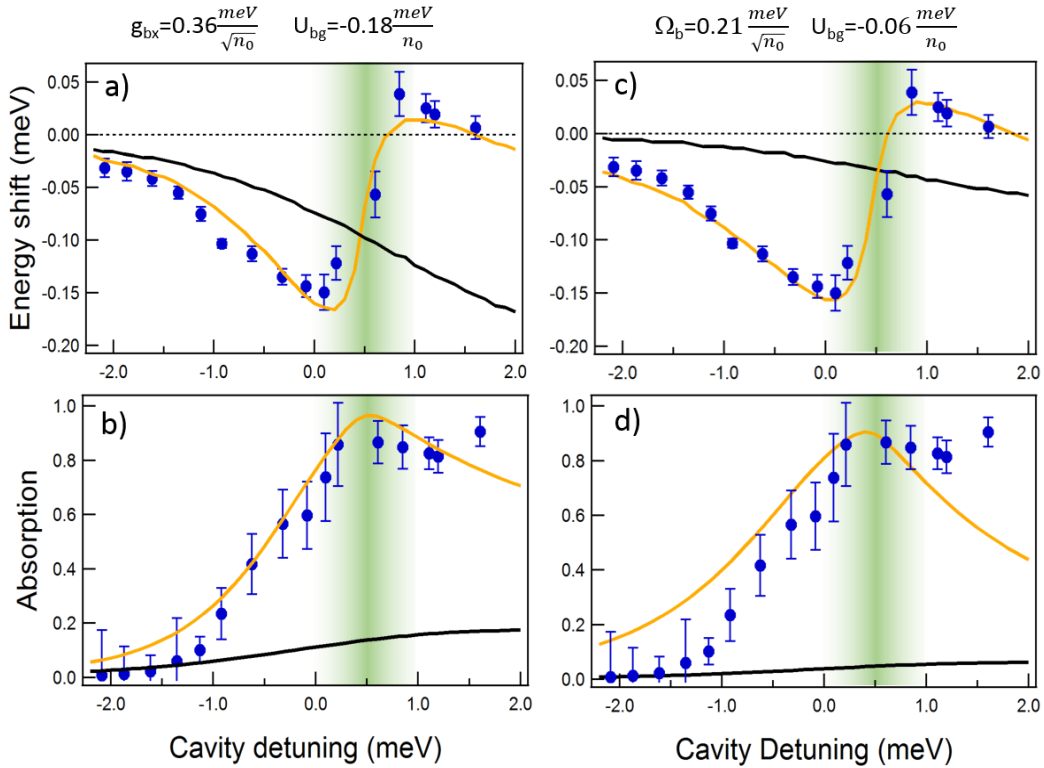

**Figure S3 Comparison between the $g_{BX}$ exciton+exciton to biexciton and $\Omega_B$ exciton+ photon to biexciton processes.** Both coupling terms reproduce the resonance features. All other parameters are same as in the paper.

In order to examine more precisely the model, we now compare our experimental results with the theoretical curves obtained by using the same value for both coupling terms $g_{BX}$ and $\Omega_B$ (Figure S4), we only change the phase term $\theta$. In Figure S4a it is clearly shown that, when setting $\theta=0$, both pathways cancel, and no resonance is observed in the theoretical results. Conversely, for phases set either to $\pi/2$ and $\pi$ the theory plots recover the Feshbach resonance (Figures S4b and c). However, again it deviates for both negative and positive detuning in the absorption feature due to the contribution of the $\Omega_B$ coupling.

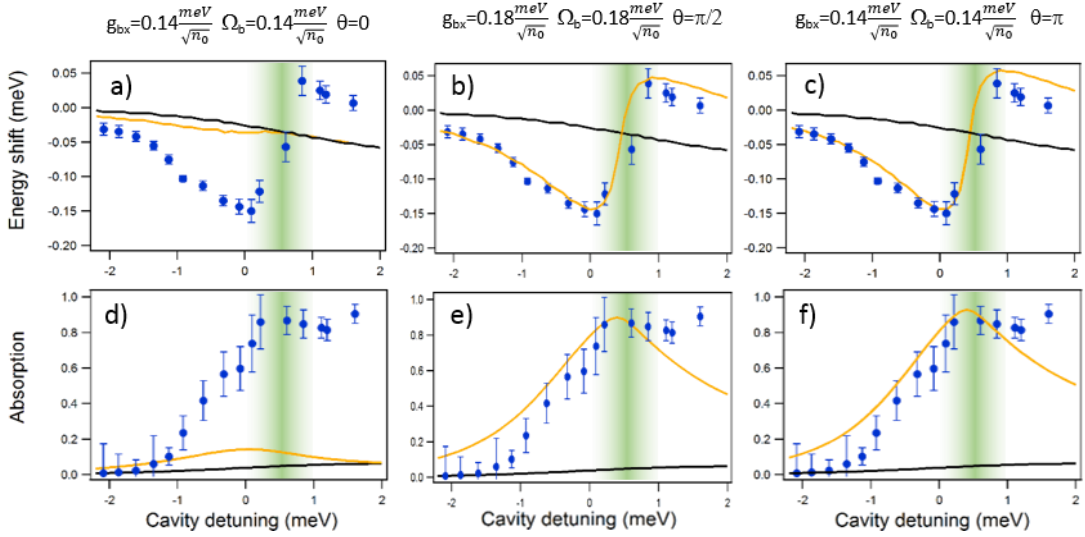

**Figure S4 Same values for the $g_{BX}$ exciton+exciton to biexciton and exciton+photon to biexciton ($\Omega_B$) processes and different phases.** The resonance feature can be reproduced for phases differing from zero. All other parameters are same as in the paper except $U_{bg}=-0.06\frac{meV}{n_0}$.

This general model enlightens that different combination of parameters and phase relations between the coupling terms are reasonably able to describe the resonance features. However, as the phase value cannot be determined a priori within this generalized model, we cannot obtain, with the limited information available, a precise determination of all parameters. In particular, each term alone gives a reasonable account of the experimental results. The lesser quality of the fit when considering only the polariton-biexciton coupling through the term $\Omega_B$, although it gives rise to a nice Feshbach resonance, brings us to favor the bipolariton model with the Coulomb term $g_{BX}$ as unique polariton-biexciton coupling. This gives us the best fit to our Feshbach resonance results.